\documentclass{aa}  

\usepackage{graphicx}
\usepackage{txfonts}
\usepackage{color}
\usepackage[colorlinks=true,urlcolor=blue,citecolor=blue,linkcolor=blue]{hyperref}
\begin{document}

   \title{On the Origin of Magnetic Perturbations associated with the FIP Effect}

   \author{M. Murabito,
          \inst{1}
          M. Stangalini,
          \inst{2}
          D. Baker,
          \inst{3}
          G. Valori, 
          \inst{4}
          D.~B. Jess, 
          \inst{5,6}
          S. Jafarzadeh,
          \inst{7,8}
          D.~H. Brooks,
          \inst{9}
          I. Ermolli,
          \inst{1}
          F. Giorgi,
          \inst{1}
          S.~D.~T. Grant,
          \inst{5}
          D.~M. Long,
          \inst{3}
          L. van Driel-Gesztelyi
          \inst{3, 10, 11}
          }

\institute{INAF Istituto Nazionale di Astrofisica, Osservatorio Astronomico di Roma, 00078 Monte Porzio Catone, RM, Italy \\
\email{mariarita.murabito@inaf.it}
\and
ASI - Agenzia Spaziale Italiana, Via del Politecnico snc, Rome, Italy 
\and
University College London, Mullard Space Science Laboratory, Holmbury St. Mary, Dorking, Surrey, RH5 6NT, UK
\and
Place 
\and
Astrophysics Research Centre, School of Mathematics and Physics, Queen’s University Belfast, Belfast, BT7 1NN, UK  
\and
Department of Physics and Astronomy, California State University Northridge, Northridge, CA 91330, USA 
\and
Rosseland Centre for Solar Physics, University of Oslo, P.O. Box 1029 Blindern, NO-0315 Oslo, Norway
\and
Institute of Theoretical Astrophysics, University of Oslo, P.O. Box 1029 Blindern, NO-0315 Oslo, Norway
\and
College of Science, George Mason University, 4400 University Drive, Fairfax, VA 22030, USA
\and
LESIA, Observatoire de Paris, Universit\'e PSL, CNRS, Sorbonne Universit\'e, Univ. Paris Diderot, Sorbonne Paris Cit\'e, 5 place Jules Janssen, 92195 Meudon, France
\and
Konkoly Observatory, Research Centre for Astronomy and Earth Sciences, Konkoly Thege \'ut 15-17., H-1121, Budapest, Hungary}

   \date{}

 
  \abstract{In \citet{Stangalini20} and \citet{Deb20}, 
magnetic oscillations were detected in the chromosphere of a large sunspot and found to be linked to the coronal locations where a First Ionization Potential (FIP) effect was observed. In an attempt to shed light onto the possible excitation mechanisms of these localized waves, we further investigate the same data by focussing on the relation between the spatial distribution of the magnetic wave power and the overall field geometry and plasma parameters obtained from multi-height spectropolarimetric non-local thermodynamic equilibrium (NLTE) inversions of IBIS data. We find that, in correspondence with the locations where the magnetic wave energy is observed at chromospheric heights, the magnetic fields have smaller scale heights, meaning faster expansions of the field lines, which ultimately results in stronger vertical density stratification and wave steepening. In addition, the acoustic spectrum of the oscillations at the locations where magnetic perturbations are observed is broader than that observed at other locations, which suggests an additional forcing driver to the $p$-modes. Analysis of the photospheric oscillations in the sunspot surroundings also reveals a broader spectrum in between the two opposite polarities of the active region (the leading spot and the trailing opposite polarity plage), and on the same side where magnetic perturbations are observed in the umbra. We suggest that strong photospheric perturbations in between the two polarities are responsible for this broader spectrum of oscillations, with respect to the $p$-mode spectrum, resulting in locally-excited acoustic waves that, after crossing the equipartition layer, located close to the umbra-penumbra boundary at photopheric heights, are converted into magnetic-like waves and steepen due to the strong density gradient. }

   \titlerunning{On the origin of magnetic perturbations at the base of FIP effect}
   \authorrunning{M. Murabito et al.}
   \maketitle
%
\begin{figure*}[!htp]
\includegraphics[scale=0.99, clip, trim=10 70 10 0]{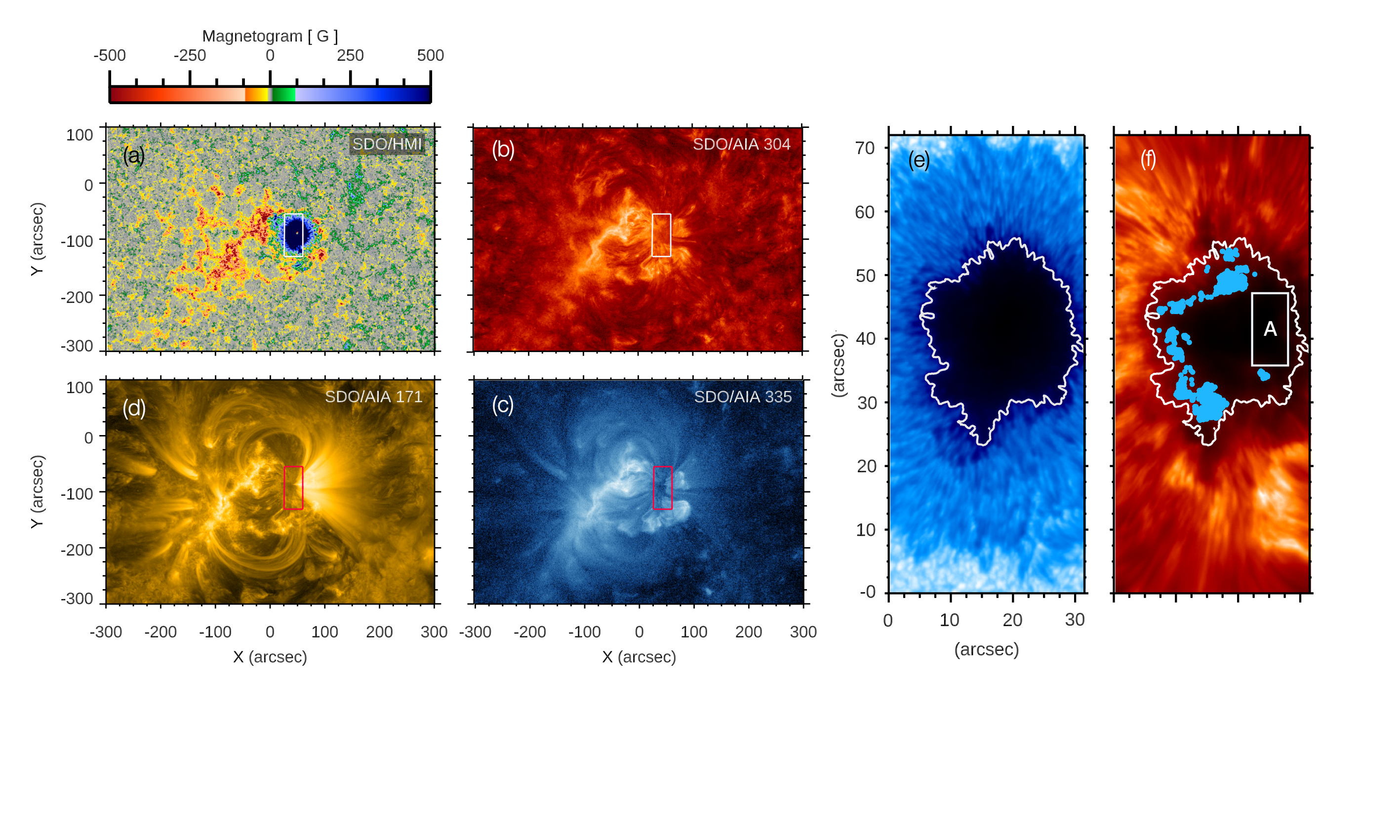}
\centering
\caption{SDO/HMI magnetogram (panel a), alongside SDO/AIA 304{\,}{\AA} (panel b), 171{\,}{\AA} (panel d), and 335{\,}{\AA} (panel c) filtergrams at the time of the IBIS observations at 14:00~UT on 2016 May 20. The white/red boxes in panels a--d indicate the IBIS FOV shown in the right panels. The HMI magnetogram is saturated at $\pm 500$~G for better visibility. Panels e and f show a photospheric continuum intensity map derived from the Fe~{\sc{i}} 617.3~nm line and a chromospheric Ca~{\sc{ii}} 854.2~nm line core intensity map, respectively. The white contours represent the umbra-penumbra boundary as derived from the continuum intensity. The blue dots indicate the locations where magnetic pertubations are detected \citep{Stangalini20}. Box `A' in the chromospheric Ca~{\sc{ii}} 854.2~nm line core image (panel~f) indicates where the probability density functions (PDFs) shown in Fig.~\ref{fig:histo} have been computed.} A movie of HMI magnetograms is available in the online material.
\label{fig:fig1_mappe}
\end{figure*}

\section{Introduction}

Although one would expect the solar corona to have the same elemental abundances of the solar photosphere, this is not always the case \citep{Pottasch1963,Meyer1985a,Meyer1985b,Widing&Feldman1989,Widing&Feldman1995,Sheeley1995,Sheeley1996}. 
The abundance variation observed in the corona depends on the first ionization potential (FIP) of an element.
Elements with FIP less than approximately 10 eV are enhanced in the corona by a factor of 3--4 compared to the photosphere whereas those elements with FIP greater than 10 eV tend to maintain their photospheric abundances.
This FIP effect is measured using the FIP bias which is the ratio of an element's abundance in the solar atmosphere to its abundance in the photosphere. Interestingly, the FIP effect is also observed in the solar wind, where it was suggested as a means to link components back to their source regions in the solar atmosphere \citep[e.g.][]{Brooks2011,Brooks2015, Hinodereview2019}. 

It is argued that the FIP effect can be due to the ponderomotive force linked to the magnetic oscillations associated with magnetohydrodynamic (MHD) waves \citep{Laming2015}.
The ponderomotive force arises from the reflection/refraction of the magnetic-like waves in the chromosphere and acts only on the low FIP ions while leaving the mainly neutral high FIP elements unaffected.
Ions are separated from neutral elements in the chromosphere and then only the ions are transported to the corona where they may be observed with enhanced abundances compared to those of the photosphere. 
However, no observational evidence of this scenario was available until very recently when, by exploiting a unique combination of high resolution observations in the chromosphere and corona with magnetic modelling, it was possible to detect magnetic perturbations in a sunspot chromosphere and find a link with the high FIP bias locations in the corona above the same sunspot \citep[][hereafter referred to as papers A and B, respectively]{Stangalini20,Deb20}. These results were also in agreement with previous studies of the same magnetic structure, where the presence of intermediate (Alfv{\'{e}}n) shocks were reported at the same locations \citep{Houston2020}. However, although providing observational support to link the FIP effect to magnetic-like waves \citep{Laming2015,Laming2017}, in papers A and B only a few possibilities were put forward to explain the surprising localised presence of magnetic perturbations only at particular locations within the sunspot umbra.

Paper A reported that the magnetic perturbations were only detected on one side of the sunspot, thus suggesting a possible role of the magnetic field geometry or the connectivity with surrounding diffuse magnetic fields. The authors suggested MHD mode conversion at the Alfv{\'{e}}n-acoustic equipartition layer (i.e, where the Alfv{\'{e}}n and acoustic speeds nearly coincide; $v_{A}=c_{s}$) as a possible cause, in agreement with \citet{Houston2020}, who detected intermediate shocks in the equipartition layer that was estimated to reside between the upper photosphere and lower chromosphere.\\
In general, waves entering the region where the Alfv{\'{e}}n and acoustic speeds nearly coincide undergo a {\it{mode conversion}}  or {\it{mode transmission}} process from one form (e.g. acoustic-like to magnetic-like wave) to another \citep{Crouch2005,Suzuki2005,Cally2015}. The term `mode conversion' refers to the situation in which a wave retains its original character (i.e., fast-to-fast or slow-to-slow), yet {\it{converts}} its general nature in the form of acoustic-to-magnetic or magnetic-to-acoustic. On the other hand, with `mode transmission' one generally refers to the situation in which the wave maintains its general nature (i.e.,`magnetic-like' wave or `acoustic-like' mode), but changes character from fast-to-slow or slow-to-fast. In all cases, the attack angle, that is the angle between the wavevector and the field lines, is the dominant factor in determining both the conversion ($C$) and transmission ($T$) coefficients \citep{Cally2001,Cally2008}, with $T + |C|=1$. In particular, the fraction of incident wave energy flux transmitted from fast to slow acoustic waves is:  
\begin{equation}
    T = e^{-\pi k h_{s} sin^{2}(\alpha)} \  
    \label{eq:eqT}
\end{equation}
where $k$ is the wavenumber, $h_{s}$ the thickness of the conversion layer, and $\alpha$ the attack angle. The coefficient $C$ is a complex energy fraction to take into account possible phase changes during the process of mode conversion \citep{Hansen2009}. It was estimated that the thickness of the conversion layer can be of the order of $200-250$ km \citep{Stangalini2011}. 
From the above equation, it is clear that the conversion $C$ is larger when the attack angle is larger. This implies that the field geometry plays a significant role in the mode conversion and therefore should be carefully taken into account as postulated by paper B.\\
In this work, in an attempt to shed light on the different mechanisms generating the FIP effect, we investigate the wave propagation across different heights above the sunspot as a function of the plasma and magnetic field parameters, as inferred from multi-height spectropolarimetric inversions. 
For this purpose, we make use of a combination of high-resolution spectropolarimetric observations acquired by IBIS in the photosphere and chromosphere, SDO/HMI line-of-sight (LOS) Dopplergrams, and SDO/AIA data to determine the wave flux across different layers of the solar atmosphere and analyze its relation to the global parameters such as inclination angles, vertical gradients of the magnetic field, and density ratios  of the magnetic region.\\
This study can be preparatory for the scientific exploitation of future space missions such as Solar-C EUVST and Solar Orbiter.

\section{Observational Data}

The dataset used in this work was acquired with the Interferometric BIdimensional Spectrometer \citep[IBIS;][]{Cav2006} instrument at the Dunn Solar Telescope (DST) on 2016 May 20 under excellent seeing conditions for more than two hours, between 13:40 -- 15:30~UT. This dataset has been the main focus of other studies \citep[see, e.g.,][]{Stangalini18,Murabito19,Houston2020,Murabito20,Stangalini20,Deb20}, due to the quality of the data and the large-scale nature of the observed sunspot, which was the leading spot of AR 12546.

The observations were carried out using the Fe~{\sc{i}} 617.3~nm and Ca~{\sc{ii}} 854.2~nm lines with a sampling of 20~m\AA~and 60~m\AA, respectively. Both lines were acquired in spectropolarimetric mode with 21 spectral points and a cadence of 48~s. A standard calibration procedure (flat field, dark subtraction, polarimetric calibration) was first applied. In order to remove the residuals of atmospheric aberrations, the dataset was processed with the Multi-Object Multi-Frame Blind Deconvolution \citep[MOMFBD;][]{vanNoort2005} technique. From the final IBIS cubes, the circular polarisation (CP) signals (for both the photospheric and chromospheric lines) were calculated pixel-by-pixel following the definition given in \citet{Stangalini20}, using the maximum amplitude of the Stokes-V spectral profile.

To complement the IBIS data and better study the wave power, we use full-disk Dopplergrams acquired by the Helioseismic and Magnetic Imager \citep[HMI;][]{Schou2012} on board the Solar Dynamics Observatory \citep[SDO;][]{Pesnell2012} satellite in the interval between 13:00 -- 16:00~UT, with a cadence of 45~s. The pixel scale of these data is 0.5\arcsec. We also analyzed simultaneous Atmospheric Imaging Assembly \citep[AIA;][]{Lemen2012} EUV filtergrams taken in the 304{\,}{\AA}, 171{\,}{\AA}, and 335{\,}{\AA} passbands. The pixel scale of the SDO/AIA data is 0.6\arcsec and the cadence is 12~s. 

The combined IBIS, SDO/HMI and SDO/AIA data are used to investigate the spatial distribution of the wave power penetrating the higher layers of the sunspot atmosphere, in order to obtain a tomographic view of the embedded MHD processes. Figure \ref{fig:fig1_mappe} shows an overview of AR~12546 as observed by the SDO/HMI and SDO/AIA EUV (304{\,}{\AA}, 171{\,}{\AA}, and 335{\,}{\AA}) instruments (panels a, b, c and d) and by the IBIS instrument (panels e and f) on 2016 May 20. The IBIS field-of-view (FOV) captured one of the biggest sunspot of cycle~, manifesting as a strong coherent leading positive polarity sunspot, as displayed in the SDO/HMI magnetogram and the photospheric IBIS continuum intensity maps in Fig.~\ref{fig:fig1_mappe} (panels a and e). The AR at the time of IBIS observations was located near the disc center, at X=35\arcsec~and Y=-90\arcsec. The magnetogram also shows an asymmetric flux distribution between the trailing and leading side of the moat region around the biggest sunspot. Indeed, moving magnetic features (MMF) activity is observed, which is asymmetric, being more extended and vigorous on the left (east) side of the umbra, coinciding with the segment where the blue dots are observed in Fig.~\ref{fig:fig1_mappe}f.

\begin{figure*}[!htp]
\includegraphics[scale=0.3, clip, trim=10 150 10 0]{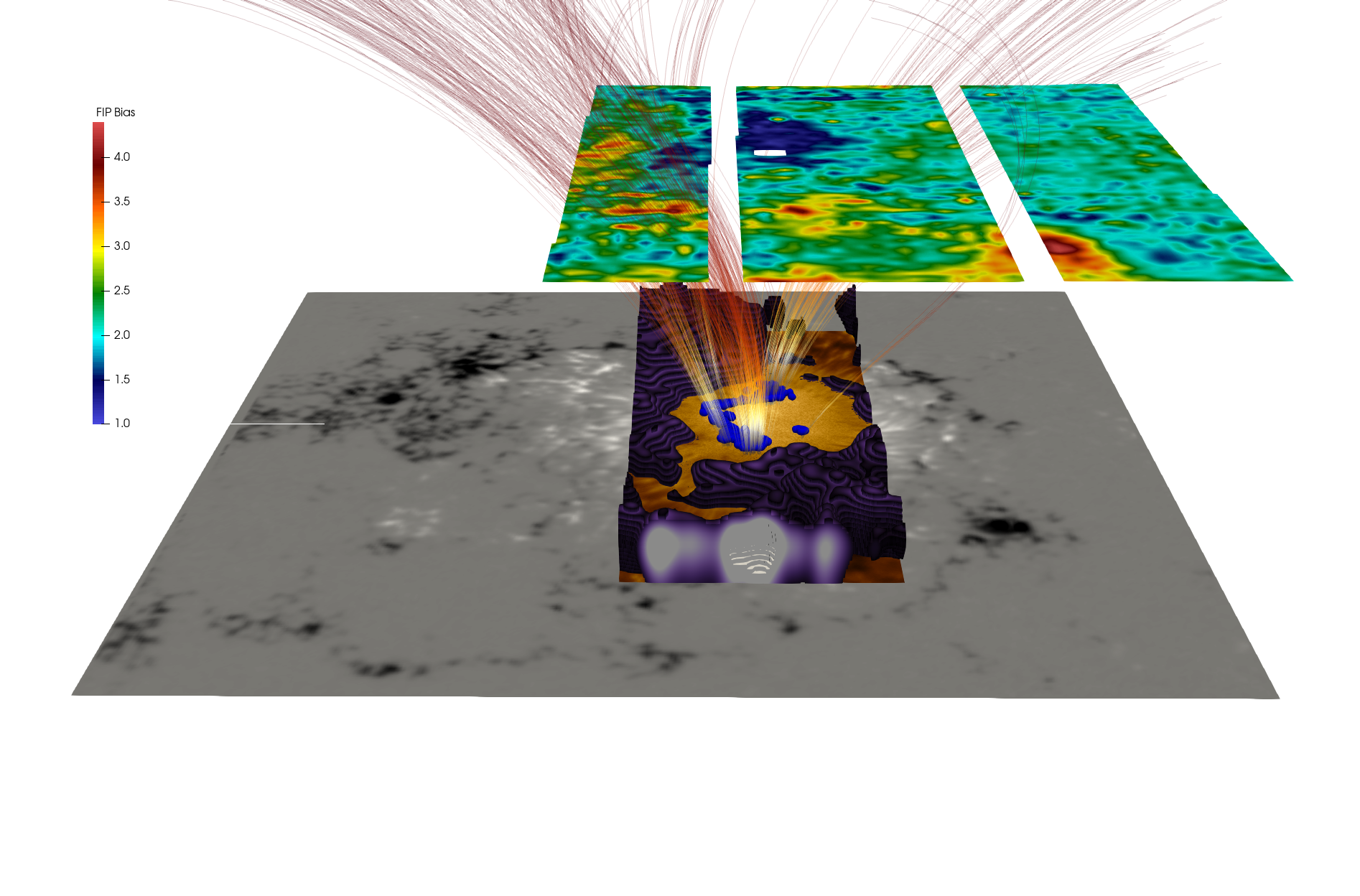}
\centering
\caption{Three-dimensional view of the AR 12546. From bottom to top: SDO/HMI magnetogram, IBIS Fe~{\sc{i}} core with blue dots overplotted, and \textit{Hinode}/EIS FIP bias map. The purple surface represents the equipartition layer ($0.8< c_{s}/v_{A} < 1.2$) as inferred from the spectropolarimetric inversions. Selected field lines from a PFSS extrapolation of the coronal field link the blue dots with regions of high FIP bias on the eastern and southern edges of the sunspot i.e. in the penumbra.   See \cite{Deb20} for more details.  
\label{fig:fig2static_view}}
\end{figure*}


\section{Methods and Results}
\subsection{Magnetic perturbations and local properties of the sunspot}

\begin{figure}[h]
\includegraphics[scale=0.48, clip,trim=170 40 50 50]{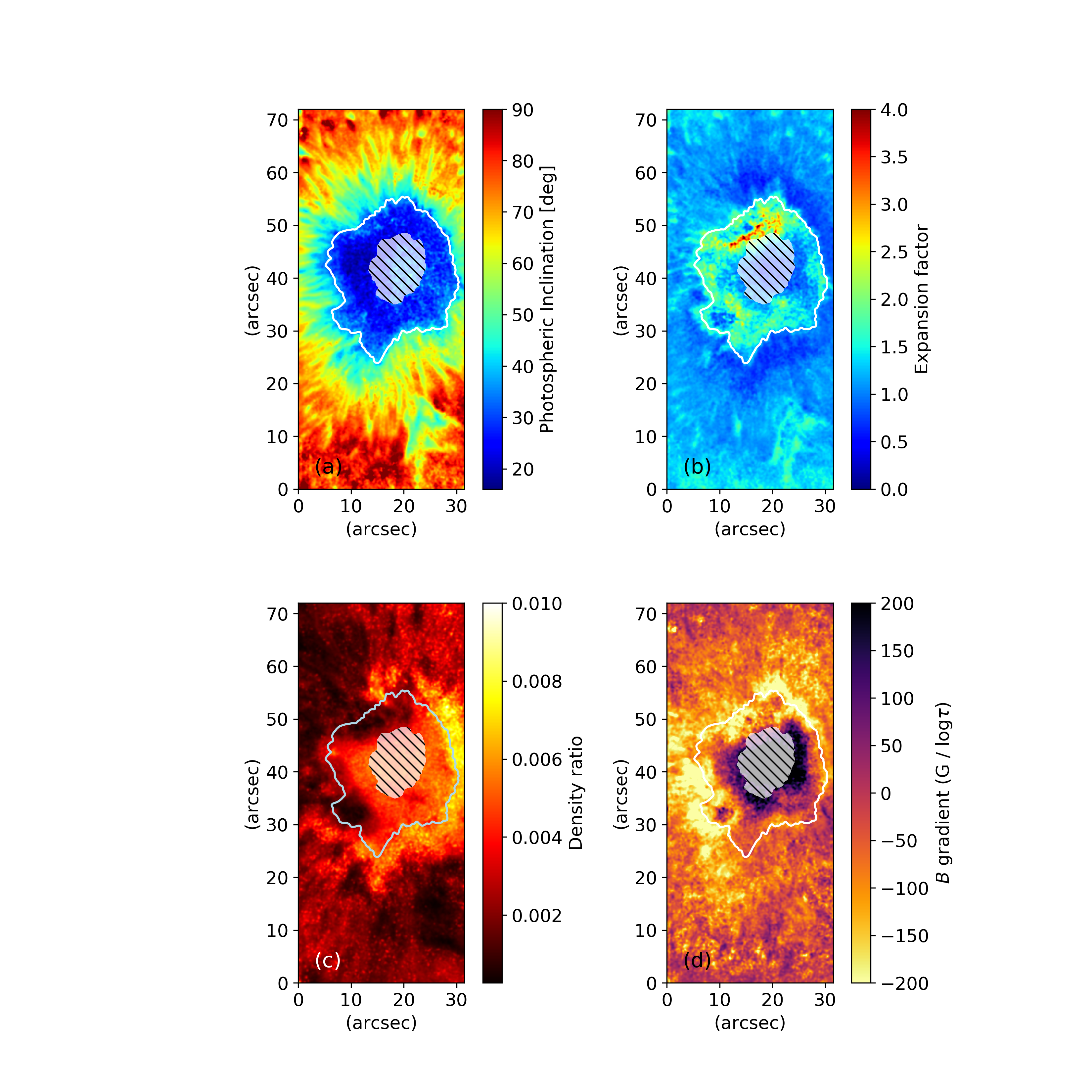}
\caption{Inclination angle of the magnetic field at $\log_{10}(\tau)=-1.0$ (photosphere; panel a). Expansion factor of the magnetic field between photosphere and chromosphere (i.e. at $\log_{10}(\tau)=-4.6$) (panel b). Density ratio between chromosphere and photosphere (panel c). Total magnetic field gradient (panel d). The white contour in all maps represents the umbra-penumbra boundary as seen in the continuum intensity. The hatched area indicates the central region of the umbra, where saturation effects and low
photon flux are detected in the photosphere \citep{Stangalini18}. The magnetic field, the inclination angles, and the density ratios were derived from the NICOLE inversions. } 
\label{fig:fig_inversion}
\end{figure}

In paper B, the authors used observations obtained with the EUV Imaging Spectrometer \citep[EIS;][]{Culhane2007} on board the \emph{Hinode} \citep{Kosugi2007} satellite to make a spatially resolved map of coronal composition, or FIP bias, in the region of the sunspot (see Fig.~\ref{fig:fig2static_view}).
Highly fractionated plasma with FIP bias of 3$^{+}$ is observed in loops rooted in the penumbra on the eastern and southern edges of the sunspot whereas the coronal field above the umbra contains unfractionated plasma (FIP bias of 1--1.5).
On the west, FIP bias is approximately 2-2.5. 

In order to investigate the role of the wave dynamics on the FIP effect observed at higher layers, we studied the spatial distribution of the wave power and compared it to the magnetic field and plasma parameters (namely, the field inclination, density ratio, and vertical gradient of the magnetic field) as inferred from spectropolarimetric inversions. The location of magnetic oscillations detected in Papers A and B are shown in Fig. ~\ref{fig:fig1_mappe} (panel f) and Fig.~\ref{fig:fig2static_view} as blue dots. These blue dots are not uniformly distributed within the umbra of the sunspot, but are only located towards the left side of it, close to the umbra-penumbra (UP) boundary, therefore on the same side as the trailing negative polarity of the AR (as shown by the magnetogram at the bottom of the three-dimensional view of Fig.~\ref{fig:fig2static_view}). 
Using a Potential Field Source Surface (PFSS) extrapolation to model the  magnetic field of the corona, the locations of the blue dots were magnetically linked to regions of high FIP bias at coronal heights as shown in Fig.~\ref{fig:fig2static_view} (see paper B for more details). 

The magnetic field geometry is examined by using the non-local thermodynamic equilibrium (NLTE) inversions already presented in \citet{Murabito19}, which were carried out using the NICOLE code \citep{Nicole2015} on the same data (i.e. the best spectral scan of the data series, in terms of contrast). Both the photospheric and chromospheric lines are inverted simultaneously, thus providing a three-dimensional stratification of the most relevant atmospheric parameters in the range of explored heights. More details on the inversion procedure can be found in \citet{Murabito19}, although here we summarize the key points for completeness. As a first step, we investigated the atmospheric parameters obtained from the spectropolarimetric inversions at the location of the blue dots (Fig.~\ref{fig:fig1_mappe}f and Fig.~\ref{fig:fig2static_view}), focusing our attention on the two atmospheric heights corresponding to the maxima of the response functions of the two spectral lines (to magnetic field perturbations), i.e., $\log_{10}(\tau)\approx-1.0$ for the photospheric Fe~{\sc{i}} line and $\log_{10}(\tau)\approx-4.6$ for the chromospheric Ca~{\sc{ii}} line as reported in \citet{Murabito19}, and in agreement with the previous study by \citet{Quintero2016}.

\begin{figure}[!t]
\centering
\includegraphics[scale=0.35, clip,trim=0 0 0 50]{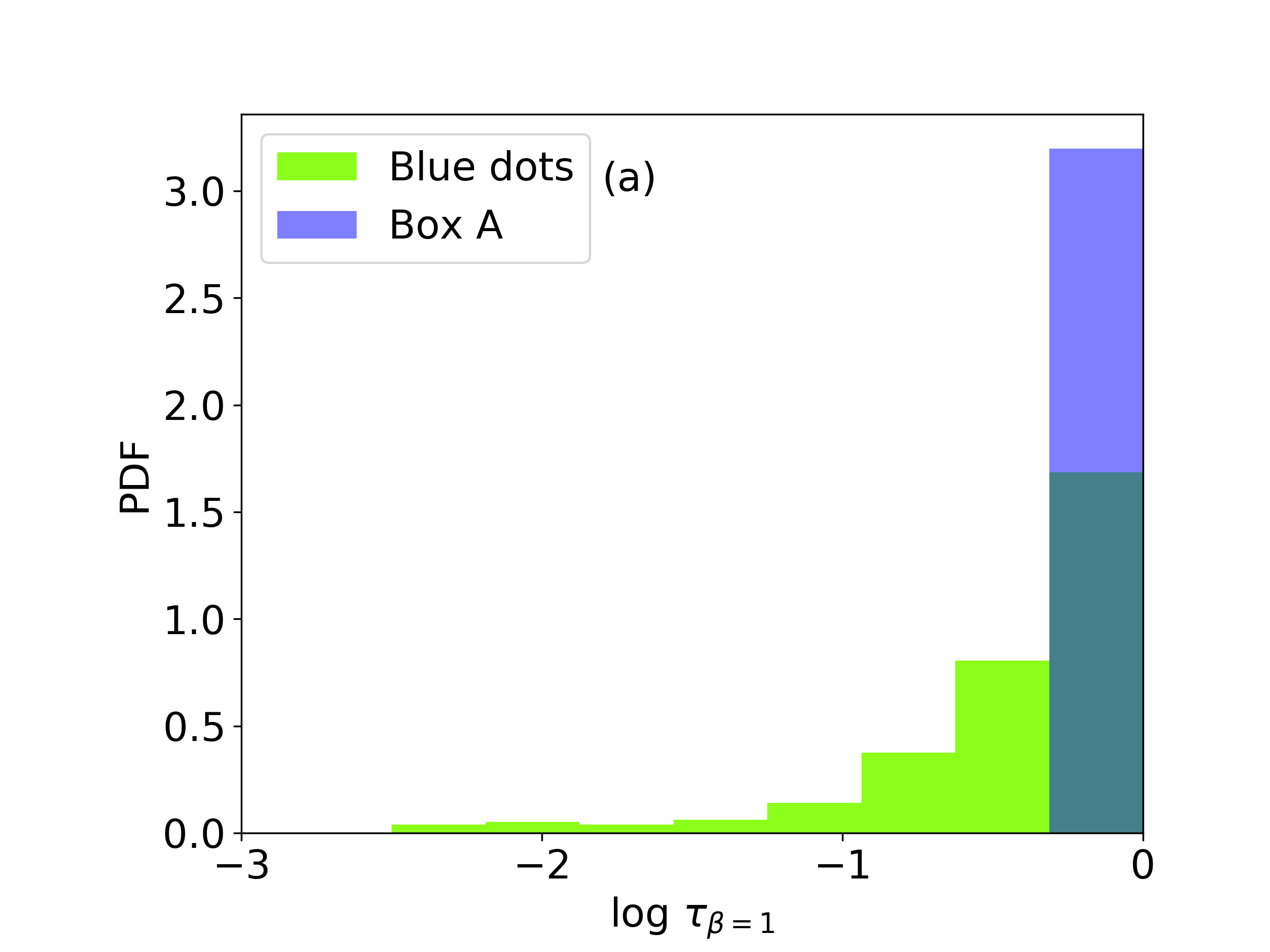}
\includegraphics[scale=0.35, clip,trim=0 0 0 50]{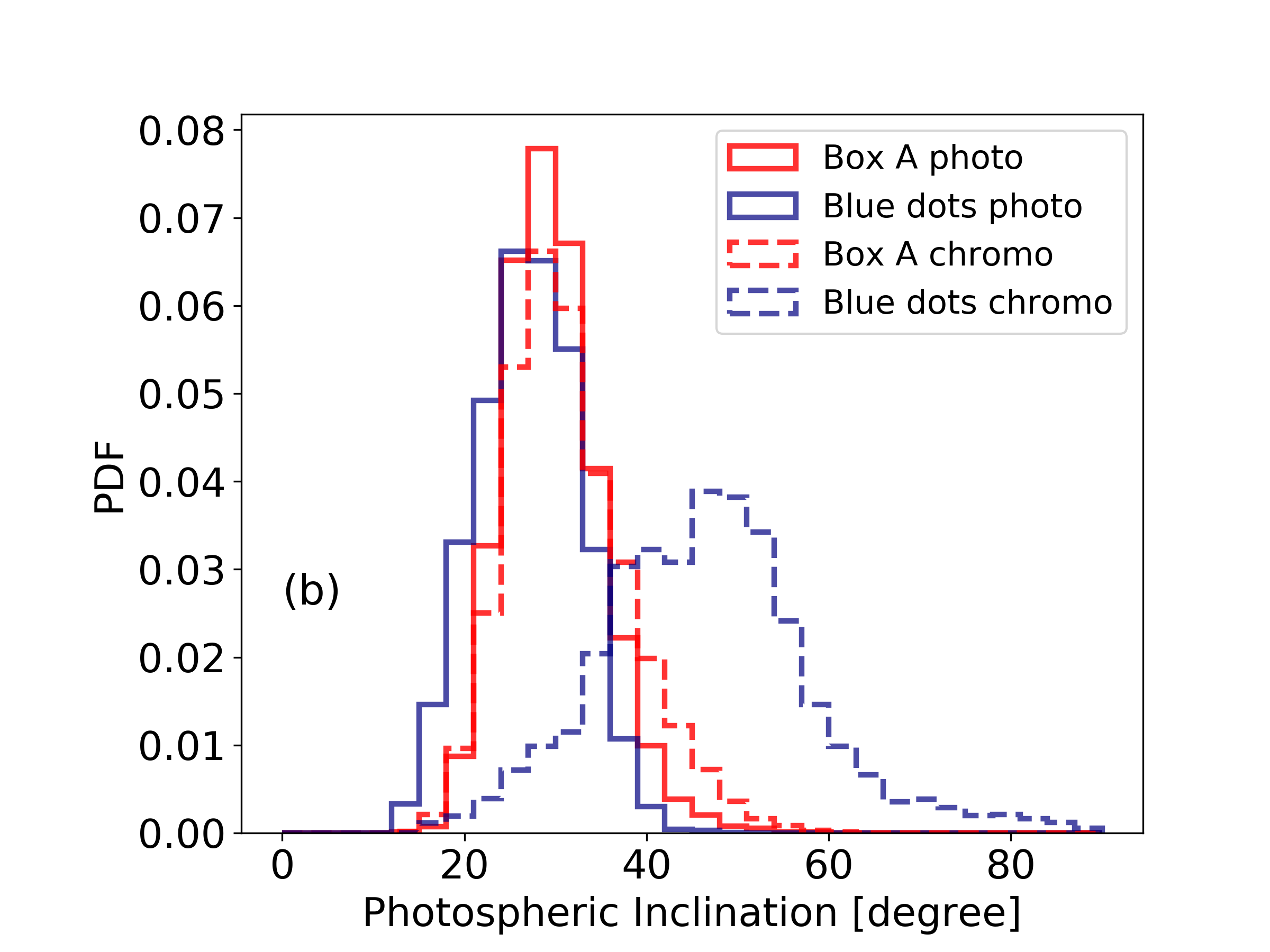}
\includegraphics[scale=0.35, clip,trim=0 0 0 40]{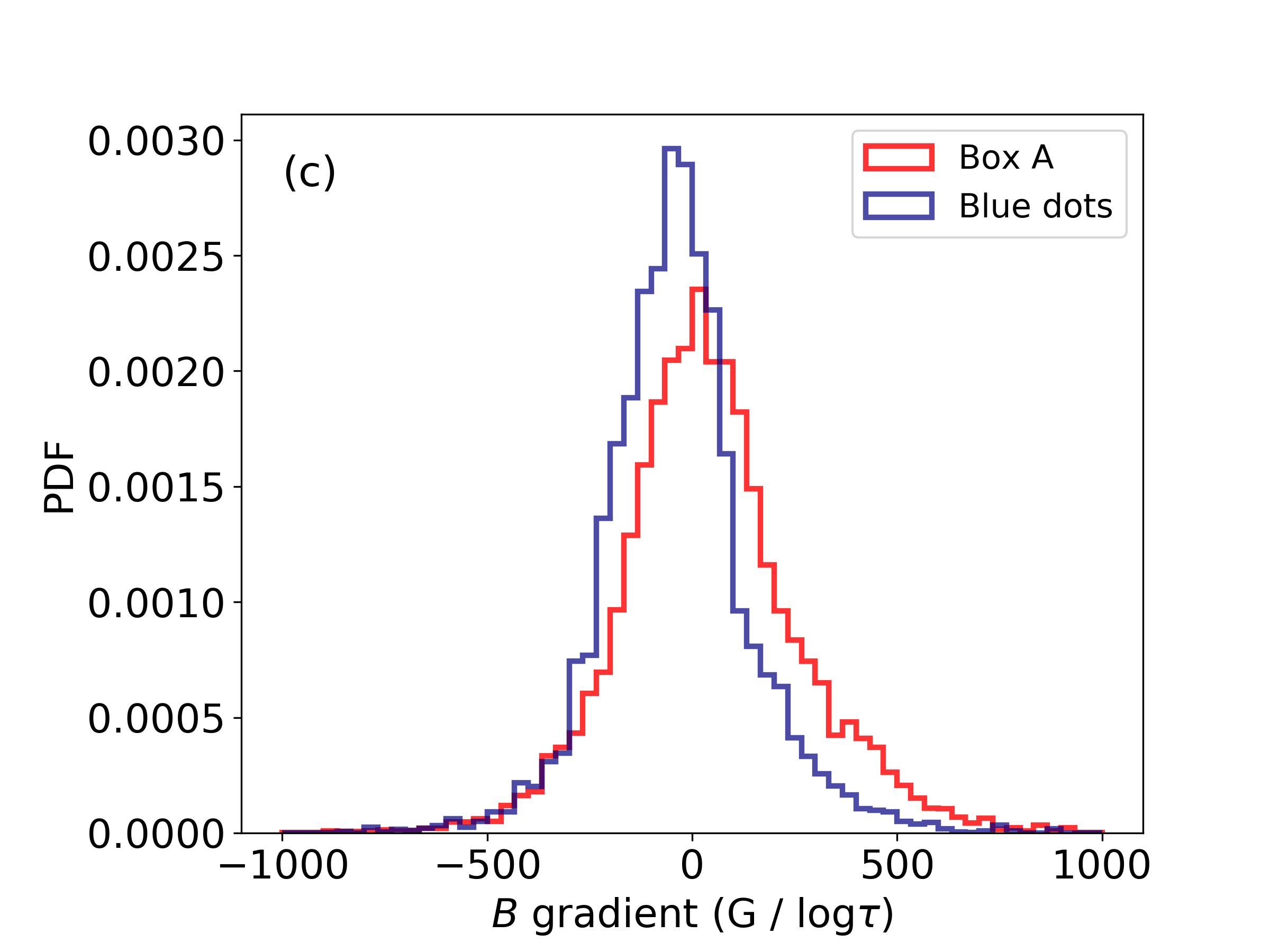}
\includegraphics[scale=0.35, clip,trim=0 0 0 40]{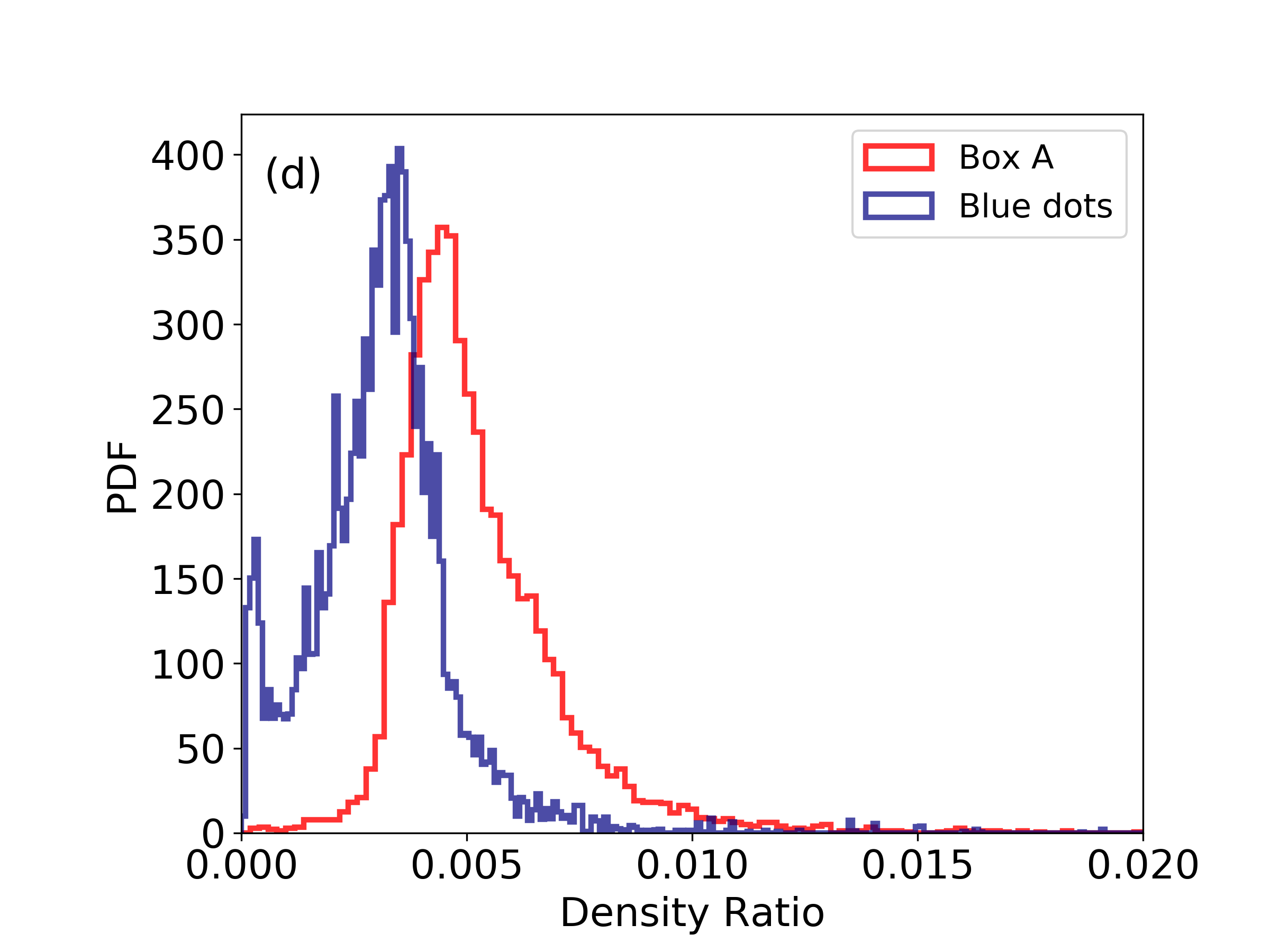}
\caption{(Panel a) PDF of the optical depths, $\log_{10}(\tau)$, corresponding to the equipartition layer ($v_{A}=c_{s}$) associated with the locations of the blue dots and the area inside box A shown in Fig.~{\ref{fig:fig1_mappe}f}. (Panel b) PDFs of the magnetic field inclination angles at photospheric ($\log_{10}(\tau)=-1.0$, red and blue histograms) and chromospheric ($\log_{10}(\tau)=-4.6$; dashed histograms) heights. (Panel c and d) PDF distribution of the magnetic field gradient and density ratio for the selected region (i.e. the blue dots and the box A).
} 
\label{fig:histo}
\end{figure}

In Fig.~\ref{fig:fig_inversion} we show the photospheric (panel a) 
magnetic field inclination maps. Given the location of the AR during the observations, i.e., at solar disc center, we neglect any projection effects. Hence, the displayed maps are consistent with the LOS reference frame. Here, we see that at photospheric heights there is no significant difference in the inclination angle (on average) between the left and right sides of the umbra. However, at chromospheric heights we note a slightly larger magnetic field inclination corresponding to the location of the blue dots (see Fig.~{\ref{fig:fig1_mappe}}f and Fig.~\ref{fig:fig2static_view}), and thus with the locations where the magnetic perturbations linked to the coronal FIP effect are detected (see the FIP bias map in Fig~\ref{fig:fig2static_view}). In order to examine the potential role of the mode conversion process, that occurs at the equipartition layer, we calculated the probability density function (PDF) of the optical depths (in $\log_{10}\tau$) corresponding to this layer (i.e., $c_{s}=v_{A}$), which is displayed in Fig.~{\ref{fig:histo}}a. This layer can play a significant role in the wave energy conversion \citep{Grant2018}, and as can be seen in Fig.~{\ref{fig:histo}}a, it is predominantly located very close to the low photosphere (i.e., $-1\geq\log_{10}(\tau)\geq0$). In particular, we can note that the right side of the umbra (i.e. blue PDF in Fig.~{\ref{fig:histo}}a) has the equipartition layer at much lower geometric heights, while on the opposite side (i.e., at the location of the blue dots, green PDF in Fig.~{\ref{fig:histo}}a) this region extends more into the mid-photosphere. 

To better investigate if the magnetic field geometry could be responsible for a possible mode conversion process, we plot the distribution of the magnetic field inclinations at photospheric heights, close to the equipartition layer, at the locations corresponding to the blue dots and an area on the opposite side (see the box A in Fig.~\ref{fig:fig1_mappe}f). The PDFs of the photospheric magnetic field inclinations (Fig.~\ref{fig:histo}b) show a similar distribution for both opposite locations in the umbra, with a field inclination centered around 25$^{\circ}$ -- 30$^{\circ}$. In any case the small differences in the two distributions, if statistically significant, can not play a role in the mode conversion process (see Eq.~\ref{eq:eqT}) and justify the presence of magnetic-like disturbances on only one side. This is true for waves travelling in all directions, that would experience a similar conversion efficiency. However, we can not rule out the possibility of an asymmetry in the acoustic driver itself. This aspect will be better addressed in Sect. 3.2. For comparison, Fig.~\ref{fig:histo}b displays the chromospheric inclination angles at the two considered sides (dashed histograms).

To better highlight the differences between the chromosphere and photosphere, we report the chromospheric magnetic field inclinations in relation to their photospheric counterparts, i.e., a measure of the expansion factor, in Fig.~\ref{fig:fig_inversion}b. This map shows that at both sides (i.e., left and right sides of the umbra) the chromospheric magnetic field is more inclined (by about a factor 1.5) but at the left side, at the location of the blue dots, this factor reaches 3 -- 4 times larger. This finding indicates that the left side of the umbra experiences an expansion of the field lines about 2 times faster than the opposite side.
\\
The difference in the magnetic field inclinations at chromospheric heights is further supported by the different magnetic field gradients obtained from the inversions. Figure~\ref{fig:fig_inversion}d displays larger (up to 400~G / $\log_{10}\tau$) negative values in the locations of the blue dots (i.e. the blue histogram in Fig.~\ref{fig:histo}c), meaning more rapid decreases of the magnetic flux with atmospheric height. In fact, the opposite side of the umbra displays lower negative values (see the red histograms in Fig.~\ref{fig:histo}c). It is worth noting that there are pixels in the right side of the umbra with a positive gradient of the magnetic field, which means that the chromospheric magnetic field is stronger than the photosphere. Although this finding may appear surprising, it can be explained by considering this area as being part of the central region where the magnetic sensitivity of the photospheric spectral line saturates due to the large magnetic fields, resulting in an underestimation of the photospheric magnetic field. On the other hand, the chromospheric Ca~{\sc{ii}} line is not saturated, hence the chromospheric magnetic field inferred seems to be larger than that of the underestimated photospheric one.

The higher negative value of the magnetic field gradient found at the location of the blue dots further supports the idea of faster magnetic field expansion in the chromosphere for the regions associated with the coronal FIP effect as already seen in Fig.~\ref{fig:fig_inversion}b (blue dots in Fig.~{\ref{fig:fig1_mappe}}f and Fig.~\ref{fig:fig2static_view}). The density ratio between the chromosphere and photosphere is shown in Fig.~\ref{fig:fig_inversion}c. In agreement with the above findings, this panel and its related histogram (Fig.~\ref{fig:histo}d), illustrate that a significant density drop between the two heights occurs at the same location as the blue dots. 


\begin{figure}[!t]
\centering 
\includegraphics[width=0.8\columnwidth, clip, trim=0 40 0 40]{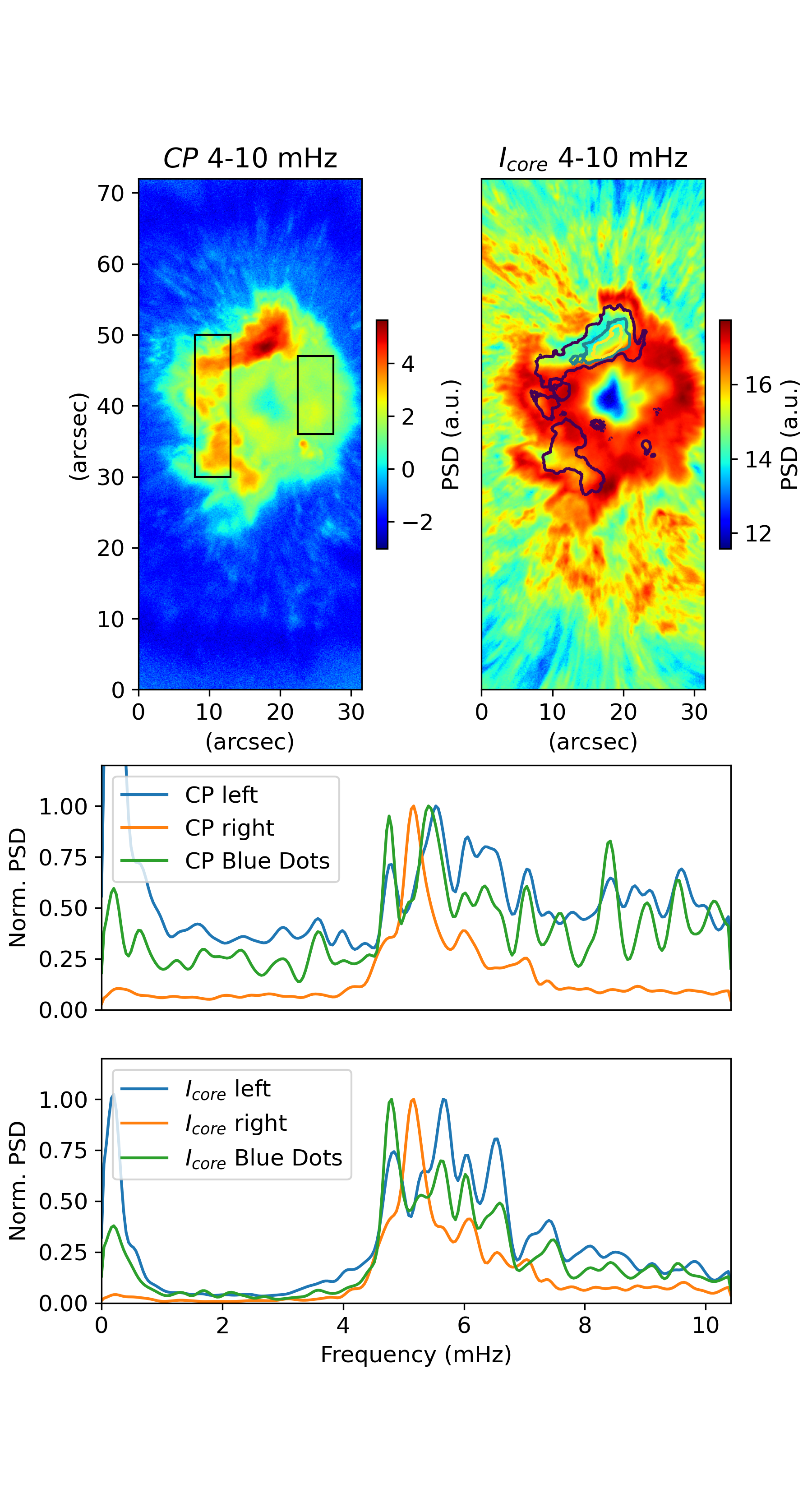}
\includegraphics[width=0.8\columnwidth, clip, trim=60 0 0 30]{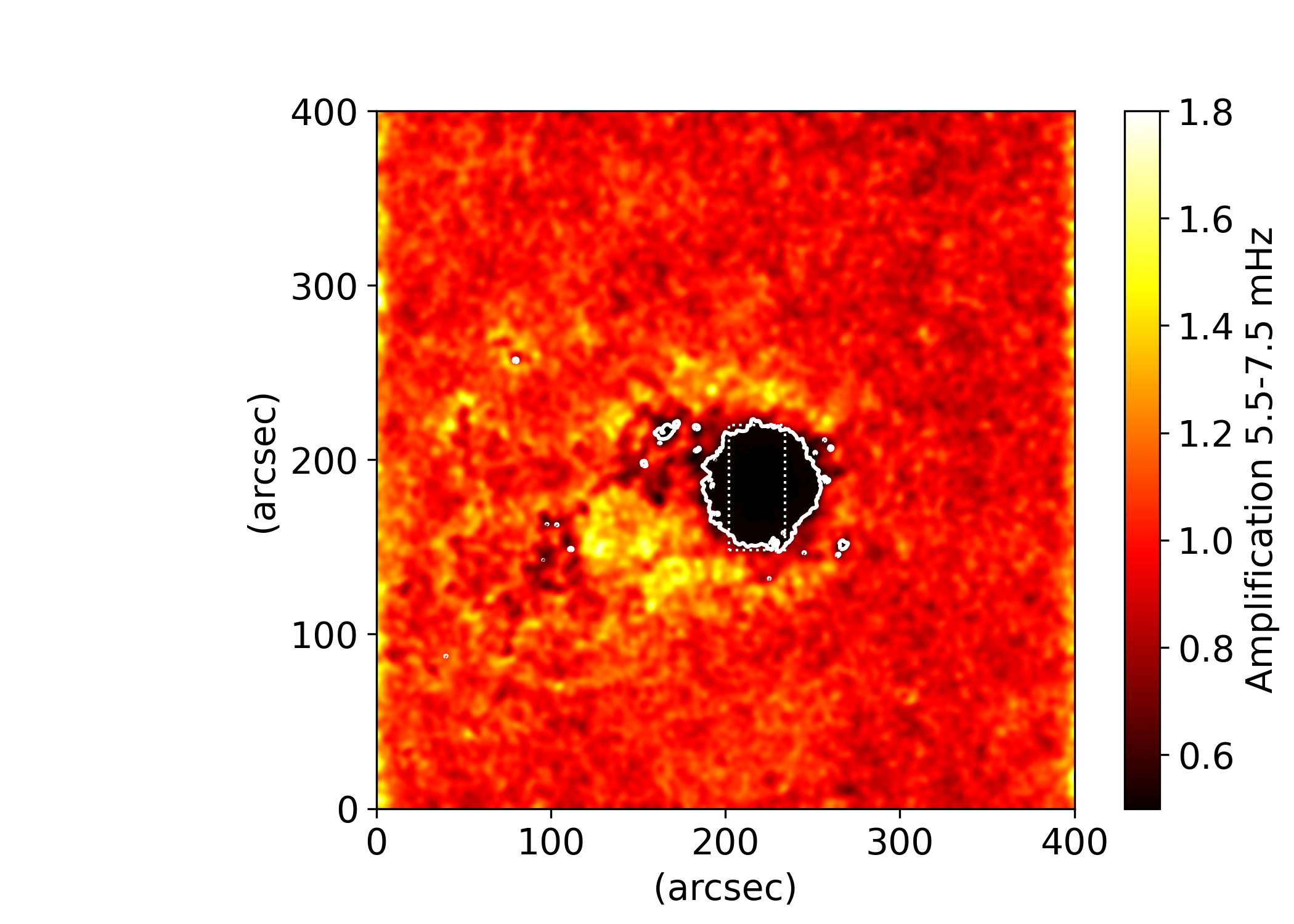}

\caption{Chromospheric IBIS CP (upper-left) and $I_{\mathrm{core}}$ (upper-right) power maps within the $4-10$~mHz range. The two boxes in the CP map indicate where the spectral averaging has been performed for deriving the results shown in the middle panels. The black contour in the $I_{\mathrm{core}}$ map represents the CP power contour where strong magnetic wave power is detected. Plots of the CP (middle-top) and $I_{\mathrm{core}}$ (middle-bottom) wave power averaged across the two boxes drawn in the upper-left panel. For completeness, we also report the spectral averaging for the blue dots locations as green plots. The SDO/HMI velocity $5.5-7.5$~mHz amplification map with respect to the quiet Sun velocity power is displayed in the lower panel. The contour represents the penumbra boundary. The dotted white box in the SDO/HMI velocity amplification map (for $5.5-7.5$~mHz) indicates the IBIS FOV. 
} 
\label{fig:ibis_power}
\end{figure}

\begin{figure}[!t]
\centering 
\includegraphics[width=\columnwidth, clip, trim=25 0 12 0]{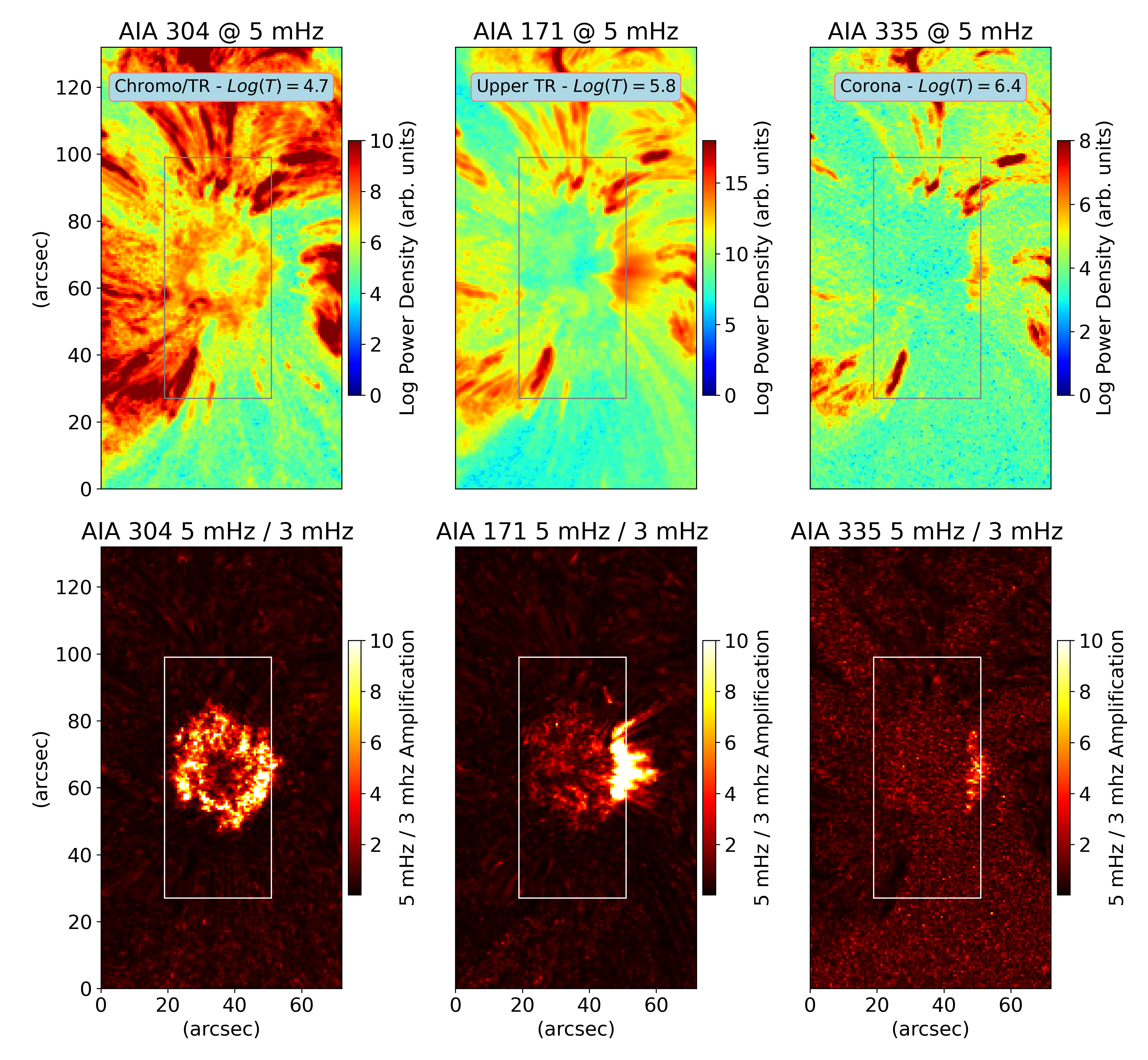}
\caption{Acoustic power in the $5$ mHz band ($1$ mHz width) for the three AIA channel spanning from upper chromosphere to the lower corona (upper row). Maps of $5$ mHz to $3$ mHz power ratio for the three AIA channels (lower row). The box in all panels indicates the IBIS FOV.} 
\label{fig:power_AIA}
\end{figure}

\subsection{Wave power as a function of height}
In addition to the analysis of the photospheric and chromospheric magnetic field geometries and their relation with the spatial distribution of the blue dots (Fig.~{\ref{fig:fig1_mappe}}f and Fig.~\ref{fig:fig2static_view}), associated with enhanced coronal FIP effects (see Fig.~\ref{fig:fig2static_view}), we have examined the spatial distribution of wave power as a function of atmospheric height. 
In particular, by employing chromospheric IBIS Ca~{\sc{ii}} circular polarisation (CP) and line core Doppler compensated intensity ($I_{\mathrm{core}}$) measurements, we computed spatially-resolved power maps averaged within the frequencies range of $4-10$~mHz, which are shown in the top panels of Fig.~{\ref{fig:ibis_power}}. This frequency range is chosen in such a way as to include the maximum of the power spectrum which is dominated by frequencies in the range of $4-4.5$~mHz in the chromosphere. These maps display unique changes in Fourier power that are cospatial with the blue dots depicted in Fig.~{\ref{fig:fig1_mappe}}f, which have been previously linked to the coronal FIP effect \citep{Deb20, Stangalini20}. 

In particular, at the locations of the blue dots, we find an excess of magnetic wave power (upper-left panel of Fig.~{\ref{fig:ibis_power}}) and a power deficit of magneto-acoustic wave power (upper-right panel of Fig.~{\ref{fig:ibis_power}}). Also, after averaging the whole spectra in two isolated boxes (towards the left and right sides of the umbra; see the boxes drawn in the upper-left panel of Fig.~{\ref{fig:ibis_power}}), it is possible to see a broader overall magnetic and magneto-acoustic power spectra (i.e., relatively large power over a wider range of frequencies) towards the left side of the umbra (middle panels of Fig.~{\ref{fig:ibis_power}}) compared to those on the right side. For completeness, we also plot the spectra at the location of the blue dots (see the green curves in the two middle panels).\\
Many authors (see for instance \citealp{Brown1992}) have reported a high frequency power enhancement in the $5.5-7.5$ mHz band around active regions at both photospheric and chromospheric heights. This effect is also referred to as acoustic halos, and is characterized by a power enhancement up to $40-60\%$ with respect to nearby quiet Sun \citep{Hindman1998}. It was suggested that this acoustic enhancement could be due to fast wave refraction due to inclined magnetic fields, in proximity of the equipartition layer \citep{Khomenko2009}.\\
In order to check whether the observed differences in the power spectrum between the left and right side of the umbra are accompanied by a similar asymmetry of the  acoustic halo, we computed the amplification map of the Doppler velocity power at high frequency ($5.5-7.5$~mHz), with respect to the quiet Sun, using SDO/HMI Dopplergrams from a larger FOV centered around the sunspot. From the lower panel of Fig.~{\ref{fig:ibis_power}}, a high frequency acoustic flux imbalance (with respect with the quiet Sun) is detected towards the left side of the sunspot, cospatial with the locations where enhanced magnetic wave activity and suppressed magneto-acoustic wave activity is found in the IBIS data sequence. The acoustic halo displays a power enhancements of the order of $50\%$, thus in agreement with previous studies \citep{Hindman1998}, although asymmetric with respect to what generally reported. \\
Furthermore, we investigated the wave power extracted from EUV intensity data associated with different temperature responses in the SDO/AIA filtergrams. In particular we have used the three SDO/AIA channels at 304{\,}{\AA} ($\log_{10}$T$ \approx 4.7$), 171{\,}{\AA} ($\log_{10}$T$ \approx 5.8$), and 335{\,}{\AA} ($\log_{10}$T$ \approx 6.4$), where T is the peak temperature response of the relevant channels \citep{Boerner2012}. As such, the selected SDO/AIA channels sample different approximate heights in the solar atmosphere, ranging from the lower transition region through to the upper corona. In particular, we look for differences in the acoustic flux transmission between these layers that can be linked to the locations where the coronal FIP effect is present (blue dots and FIP bias map shown in Fig.~{\ref{fig:fig2static_view}}). In the upper panels of Fig.~\ref{fig:power_AIA} we show the spatially-resolved power maps for the three SDO/AIA channels at a frequency of 5~mHz, which is the same frequency that dominates the chromospheric locations corresponding to the blue dots in Fig.~{\ref{fig:fig1_mappe}}f. In addition, the lower panels of Fig.~{\ref{fig:power_AIA}} show the ratios of oscillatory power between the 5~mHz and 3~mHz frequencies for each channels. Once again, despite the apparent visual symmetry of the sunspot (see, e.g., Fig.~{\ref{fig:fig1_mappe}}e), we observe an acoustic flux imbalance between the left and right side of the magnetic structure. Indeed, for the coronal channels (171{\,}{\AA} and 335{\,}{\AA}), we note that there exists specific spatial locations, notably towards the opposite side of the blue dots locations, where there is a 5~mHz power excess, while this is not the case for the 304{\,}{\AA} map, which is formed lower in the solar atmosphere and displays an almost homogeneous ring of oscillatory power \citep[similar to that shown in comparable upper-chromospheric circular-shaped sunspot umbrae;][]{Jess2013}. 

Similarly, the lower panels of Fig.~\ref{fig:power_AIA} highlight a power deficit at coronal heights cospatial with the locations displaying the enhanced FIP effect. In particular, we observe that the high-frequency acoustic flux is not able to penetrate the upper layers of the solar atmosphere at all locations, manifesting as a clear deficit of power in the location of the blue dots (shown in Fig.~{\ref{fig:fig1_mappe}}f and Fig.~{\ref{fig:fig2static_view}}). This means that the wave energy is blocked/lost at some point in the lower atmosphere, suggesting that the umbral magneto-acoustic waves linked with the blue dots in Fig.~{\ref{fig:fig1_mappe}}f are unable to reach coronal heights. 
It is important to keep in mind that the SDO/AIA filtergrams are sensitive only to intensity (proxy for density in optically thin media) fluctuations and it is possible that these fluctuations exist but they are smaller than the detection limit of the instrument. 

We summarize our main observational findings as follows:
\begin{enumerate}
    \item The averaged \textit{photospheric} inclination angle of the magnetic field lines are not significantly different between the left and right sides of the umbra
    \item Despite the apparent symmetry of the umbra, at the locations corresponding to the blue dots in Fig.~{\ref{fig:fig1_mappe}}f (i.e., the locations linked with the coronal FIP effect) we measure a faster expansion of the magnetic field lines and a significant drop in vertical plasma density. 
    \item High-frequency (i.e., $4-10$~mHz) power maps of chromospheric IBIS CP and $I_{\mathrm{core}}$ measurements display both an excess of magnetic power and a deficit of magneto-acoustic wave power in the locations of the blue dots in Fig.~{\ref{fig:fig1_mappe}}f (i.e., towards the left side of the umbra). This region also displays broadened CP and $I_{\mathrm{core}}$ power spectra.
    \item An asymmetric  wave flux excess at high frequencies (i.e., acoustic halo $5.5-7.5$~mHz) is observed in the photospheric SDO/HMI LOS velocity data.
    \item The acoustic flux (at 5 mHz) on the left side of the sunspot (i.e., cospatial with the blue dots depicted in Fig.~{\ref{fig:fig1_mappe}}f) does not efficiently reach coronal heights, suggesting the presence of different wave propagation mechanisms at opposite sides of the sunspot umbra. 
\end{enumerate}

\section{Discussion and Conclusions}
In papers A and B, it was found that regions of high FIP bias at coronal heights are magnetically linked to chromospheric locations where magnetic fluctuations are detected. The link between magnetic perturbations and high FIP bias was already proposed by \citet{Laming2015} and \citet{Deb20} and the aforementioned works represent observational evidence of this. Also, in papers A and B a surprising asymmetry in the spatial distribution of magnetic oscillations was observed, despite the apparent circular symmetry of the sunspot investigated. These magnetic perturbations at chromospheric heights were found to be linked to the high FIP bias locations in the corona. It was also noted that the magnetic perturbations were all located on the same side of the trailing magnetic polarity of AR NOAA 12546. In addition, from the analysis of the same target at chromospheric height, \citet{Houston2020} observed the signatures of intermediate shocks. Although papers A and B were not focused on the investigation of the spatially asymmetric wave power, but only on the detection of the magnetic perturbations and their link with high FIP bias locations in the corona, a few options were put forward as a possible explanation for the excitation of the magnetic perturbations themselves. It was speculated that the magnetic field geometry or connectivity with the outside diffuse fields could play a role in the wave excitation and propagation. Along the same line, \citet{Houston2020} also argued that the increased level of wave activity found in the same region of the sunspot could be due to the magnetic field geometry, through the mode conversion mechanism which converts MHD waves in different modes with an efficiency depending on the magnetic field inclination with respect the wavevector \citep{Schunker2006}. This point was also further commented in papers A and B. Indeed, the authors in paper A concluded that a possible role between the mode conversion and the detected magnetic disturbances could exist if one takes into account the asymmetry of the distribution of the magnetic fluctuations. In this regard, although this sunspot with its peculiarities is but one case study, the results reported here suggest a likely plausible explanation.\\

\begin{figure}[!t]
\centering 
\includegraphics[width=\columnwidth, clip]{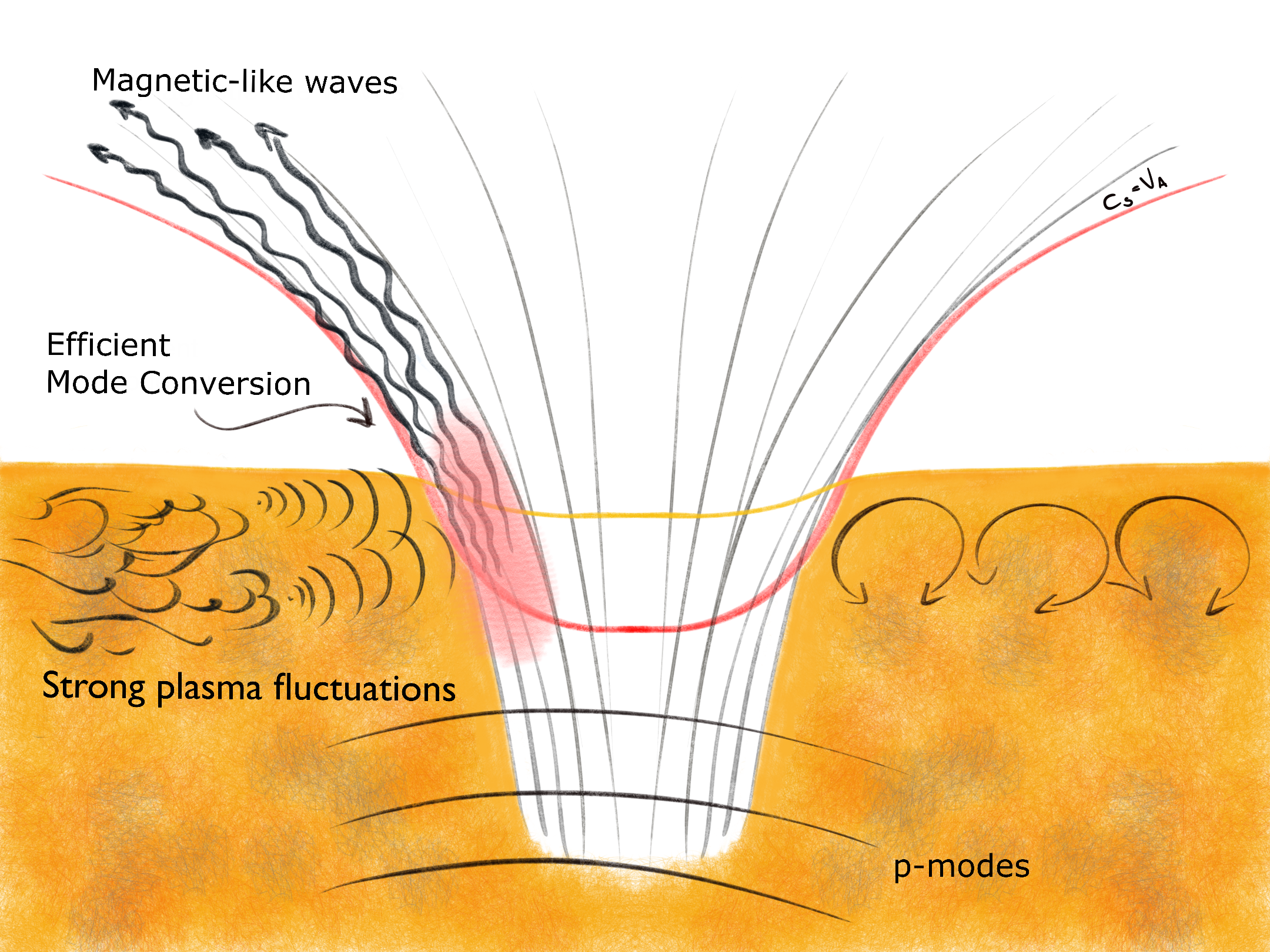}
\caption{Cartoon of the excitation mechanism of the magnetic-like waves.} 
\label{fig:Cartoon}
\end{figure}

Our results show that the magnetic field inclination at the equipartition layer (i.e. where the mode conversion takes place), located very close to the low photosphere (i.e., $-1\geq\log_{10}(\tau)\geq0$), is not significantly different on the two  opposite sides of the sunspot. This means that waves travelling in all directions (e.g. $p$-modes) would experience the same conditions (i.e. attack angle) and therefore their conversion can not justify the asymmetric wave power observed on the left side of the sunspot (i.e. blue dots). While on the one hand, we can rule out the mode conversion for waves travelling in all directions ($p$-modes), we could hypothesize the existence of a driver that acts spatially in a different way. Indeed, the study of the power spectrum at high frequencies for the chromospheric high-resolution CP signal and core intensity from IBIS data suggests to us an asymmetric photospheric driver. Furthermore, the analysis of the lower synoptic SDO/HMI observations shows an imbalanced LOS velocity power flux detected in the left side of the whole AR. Indeed, this map has shown 
a broadening of the power spectrum, referred as acoustic halo, with respect to that of a quiet sun area (where the contribution comes from the $p$-modes only). This broadening is also observed inside the umbra at the locations of the blue dots.\\
It was suggested that the halos are due to fast wave refraction in proximity to the equipartition layer and inclined magnetic fields \citep{Khomenko2009}. However, in contrast to what previously reported, in this case the halo is asymmetric and cospatial with the locations of the blue dots inside the umbra.\\
An excess of acoustic wave power in between the two polarities, due to strong photospheric plasma fluctuations in between the two polarities, could result in an excess of incident wave power on the left side of the sunspot itself (see Fig. \ref{fig:Cartoon}). In this regard, it is interesting to note that an excess moving magnetic feature activity in  between the two polarities is seen in HMI imagery covering the same observing window and depicted in Fig. \ref{fig:Cartoon} (see also online movie). This provides further evidence of an high level of small scale plasma dynamics and perturbations that may result in a variation of the acoustic field and power. We recall that MMFs are manifestations of sunspot decay and result from erosion of the sunspot’s magnetic field by turbulent plasma motions \citep[e.g.,][]{Solanki2003}. These waves travelling toward the umbra could experience a largely inclined equipartition layer before entering the umbra as shown by the purple surface representing the equipartition layer in Fig.~\ref{fig:fig2static_view}. In other words, the impact angle at the equipartition layer is particularly large, and this is an ideal condition for their  conversion into magnetic-like waves. In Fig.~\ref{fig:fig2static_view} we also note that the blue dots correspond to regions within the umbra which a wave travelling from the left side would reach immediately after crossing the equipartition layer. In addition, the equipartition layer itself, appears asymmetric with respect to the centre of the umbra (i.e. a steeper increase on the left side with respect to the right), thus possibly affecting the conversion coefficient.  

However, it is important to note that the computation of the equipartition layer requires the knowledge of gas density (to estimate the Alfv{\'{e}}n speed). The NICOLE inversion code assumes hydrostatic equilibrium for the calculation of this parameter, and this may slightly shift the inferred position of the equipartition layer. However, we would like to point out that the most relevant aspect of our interpretation is that the equipartition layer crosses and intercepts the photosphere around the umbra, not its exact position. This is guaranteed by the extreme magnetic flux of the studied sunspot. We argue that, if the above mode conversion scenario is true, the FIP effect would be aligned with the regions outside the sunspot where there is a local increase of the acoustic power. 
This fact could be directly checked in the Hinode/EIS data or, in the future, in the Solar Orbiter SPICE and Solar-C observations.

However, an additional aspect should be considered. The spectropolarimetric inversions also show a very interesting and unnoticed aspect. On the left side of the sunspot, i.e. where the magnetic perturbations were detected, the field lines undergo a faster expansion with height. As a natural result, this is also accompanied by a stronger decrease of the plasma density with height. We note that these two aspects determine the ideal conditions for the development of magneto-acoustic shocks at low heights in the atmosphere, through which the energy contained into upward propagating waves can be dissipated well before reaching the temperature minimum, as one would expect. In particular, the fast density decrease with height determines the ideal condition for a fast steepening of the wave amplitude. The intermediate shocks reported by \citet{Houston2020} and the intense magneto-acoustic activity and magnetic perturbations reported in papers A and B appear in agreement with this scenario. This is also independently corroborated by SDO/AIA observations indicating that, compared to the right side of the sunspot, acoustic wave energy on the left side is converted/dissipated much lower in the atmosphere. We argue that the large observed density drop could be responsible for the formation of shocks at very low altitudes. However, we note that the larger field inclinations found in the chromosphere towards the left side of the sunspot may provide the necessary conditions to support a high-frequency power halo outside the umbra, similar to that first reported by \citet{Toner1993} and \citet{Braun2009}. As revealed by \citet{Khomenko2009}, such power halos may be the result of wave refraction in the vicinity of the plasma-$\beta=1$ layer. Inspection of Figures~{\ref{fig:fig2static_view}} \& {\ref{fig:fig_inversion}}(d) substantiate this hypothesis, whereby the plasma characteristics present towards the left side of the sunspot act to support wave refraction, mode conversion, and ultimate dissipation of wave energy through shock formation.

Nevertheless, the question remains as to whether the faster expansion of the field lines on the left side is a consequence of the wave energy dissipation at low heights through the aforementioned shocks, or a consequence of the overall magnetic connectivity of the sunspot. Previous studies would suggest it is the latter, as the effect of shocks on magnetic field geometry is localised and any perturbations to the field are temporary as the field relaxes back to equilibrium (e.g.,  \citealp{delacruz2013, Houston2018, Bose2019}). Further, only the intermediate shocks in \citet{Houston2020} are localised to the FIP region. They detect the intensity signatures associated with slow magneto-acoustic shocks across the entire outer umbra, inferring that if shocks were capable of a bulk change in the umbral field, this would be evident on the side of the sunspot unconnected with the FIP region. 

In this regard, it is worth underlining that the context magnetogram obtained by SDO/HMI over a larger FOV shows the presence of small-scale magnetic field of opposite polarity on the same side, which constitute the trailing polarity of this AR. We argue that due to the magnetic connectivity the bending of the field lines is larger with the creation of low lying loops. This is in line with the simulations reported in \citet{Dahlburg2016}, which show that the FIP effect is stronger in short, high temperature loops. 

In addition, it is also unclear whether or not the wave steepening and mode conversion are concurrent processes. In this regard, it is worth noting that the absence of intensity fluctuations in correspondence  with the magnetic power seems to suggest that at those locations acoustic energy is mostly converted into magnetic-like waves. Further studies are necessary in order to shed light onto this aspect. 

Further, we would like to note that this kind of study has additional important implications. The FIP effect could be regarded as a proxy for the wave energy dissipation. This is a long-debated problem \citep{Jess2015} and still the identification of the physical processes with which energy is transferred from waves to the plasma remains a challenge.

Finally, our results highlight the importance of studying simultaneously different heights of the solar atmosphere, by combining simultaneous or nearly simultaneous ground-based spectropolarimetric observations in the lower atmosphere, with high spectral resolution data of the corona acquired from space. This is important not only for the investigation of the FIP effect, but also for the wave dissipation mechanisms for which the FIP effect itself could be considered as a proxy.

It is worth noting that this possibility will be widened by the new solar mission Solar-C EUVST \citep{Shimizu2011,Suematsu2016}, which will provide an unprecedented view of the corona at high temporal, spatial and spectral resolution. In this regard, this case study can be considered as a pathfinder for the full exploitation of its data in combination with high resolution spectropolarimetric imaging of the lower atmosphere.

\begin{acknowledgements}
The authors are grateful to the anonymous referee for useful comments.
This research received funding from the European Union’s Horizon 2020 Research and Innovation
531 program under grant agreements No 824135 (SOLARNET) and No 729500 (PRE-EST). This work was supported by the Italian MIUR-PRIN grant 2017 "Circumterrestrial environment: Impact of Sun-Earth Interaction" and by the Istituto NAzionale di Astrofisica (INAF).
DBJ and SDTG wish to thank Invest NI and Randox Laboratories Ltd. for the award of a Research \& Development Grant (059RDEN-1), in addition to the UK Science and Technology Facilities Council (STFC) for the consolidated grant ST/T00021X/1.
SJ acknowledges support from the European Research Council under the European Union Horizon 2020 research and innovation program (grant agreement No. 682462) and from the Research Council of Norway through its Centres of Excellence scheme (project No. 262622).
The authors wish to acknowledge scientific discussions with the Waves in the Lower Solar Atmosphere (WaLSA; \href{www.WaLSA.team}{www.WaLSA.team}) team, which is supported by the Research Council of Norway (project number 262622), and The Royal Society through the award of funding to host the Theo Murphy Discussion Meeting ``High-resolution wave dynamics in the lower solar atmosphere'' (grant Hooke18b/SCTM).
D.B. is funded under STFC consolidated grant number ST/S000240/1 and L.v.D.G. is partially funded under the same grant.
The work of D.H.B. was performed under contract to the Naval Research Laboratory and was funded by the NASA Hinode program. 
D.M.L. is grateful to the Science Technology and Facilities Council for the award of an Ernest Rutherford Fellowship (ST/R003246/1).
The italian scientific contribution to Solar-C is supported by the Italian Space Agency (ASI) under contract to the co-financing National Institute for Astrophysics (INAF) Accordo ASI-INAF 2021-12-HH.0 "Missione Solar-C EUVST - Supporto scientifico di Fase B/C/D".
\end{acknowledgements}

%
%

\end{document}